\documentclass[natbib]{svjour3}                     
\smartqed  
\usepackage{graphicx}
 \usepackage{aps-bibstyle}  
\newcommand{\aap}{{Astron. Astrophys.}}
\newcommand{\apj}{{Astrophys. J.}}

\begin{document}

\title{Plasma Diagnostics of the Interstellar Medium with Radio Astronomy
}


\author{Marijke Haverkorn         \and
        Steven R. Spangler 
}


\institute{M. Haverkorn \at
              Department of Astrophysics/IMAPP, Radboud University Nijmegen, P.O. Box 9010, 6500 GL Nijmegen, The
Netherlands\\  
Leiden Observatory, Leiden University, P.O. Box 9513, 2300 RA Leiden, The Netherlands\\
              Tel.: +31-24-365 2809\\
              \email{m.haverkorn@astro.ru.nl}           
           \and
           S. Spangler \at
              Department of Physics and Astronomy, University of Iowa\\
              Tel.: +1-319-335-1948\\
              \email{steven-spangler@uiowa.edu}
              }

\date{Received: August 02, 2013 / Accepted: ------}

\maketitle

\begin{abstract}
We discuss the degree to which radio propagation measurements diagnose conditions in the ionized gas of the interstellar medium (ISM). The ``signal generators'' of the radio waves of interest are extragalactic radio sources (quasars and radio galaxies), as well as Galactic sources, primarily pulsars. The polarized synchrotron radiation of the Galactic non-thermal radiation also serves to probe the ISM, including space between the emitting regions and the solar system.  Radio propagation measurements provide unique information on turbulence in the ISM as well as the mean plasma properties such as density and magnetic field strength.  Radio propagation observations can provide input to the major contemporary questions on the nature of ISM turbulence, such as its dissipation mechanisms and the processes responsible for generating the turbulence on large spatial scales.  Measurements of the large scale Galactic magnetic field via Faraday rotation provide unique observational input to theories of the generation of the Galactic field.   
\end{abstract}

\section{Introduction}
The purpose of this article is to discuss how radio propagation measurements provide diagnostics of the interstellar medium (ISM).  By radio propagation measurements, we mean those in which a radio astronomical observable (such as the interferometric visibility, or the polarization position angle) has been modified by a medium between the source of the radio waves and the radio telescope.  In this paper, we will be interested in plasma media.  These measurements provide rather direct information on the ionized gas density (strictly speaking, the electron density), the interstellar magnetic field, and (sometimes) indirect information on flow velocities in the interstellar medium. 

In addition to information on the mean plasma properties of the interstellar medium such as $\langle n_e \rangle$ and $\langle \vec{B} \rangle$, these propagation observations yield information on turbulent fluctuations in the interstellar plasma.  In fact, it can be argued that the information on interstellar plasma turbulence is the most unique contribution of this type of observation to studies of the ISM.  

This paper is intended, in part, to serve a tutorial and review function.  However, there have been numerous reviews in the past on the probing of the interstellar medium by radio propagation measurements, and the implications of those measurements for the astrophysics of the ISM \citep[see, in particular,][]{Uscinski77,Rickett77,Rickett90}. There is no point in repeating the material already published in those papers.  In the present article, we will make detailed reference to those papers to make a number of important points.  At the same time, we will stress remaining, open questions about the interstellar plasma and its turbulence.  In some cases, those questions have been actively discussed for many years.  We will also clearly point out and discuss those topics in which radio propagation measurements provide crucial data for some of the issues of greatest importance in contemporary astrophysics.  

An underlying theme of this paper will be the conceptual unity of plasma processes that occur in the interstellar medium, the solar corona, the interplanetary medium, and finally, experiments in plasma physics laboratories. In the last decade or two, plasma physics laboratory experiments have succeeded in illuminating processes which also occur in astrophysical plasmas.  These experiments deal with processes which are at the basis of plasma astrophysics.  A partial list of the experiments which are contributing a new dimension to plasma astrophysics are measurement of Faraday rotation in laboratory plasmas, and its use in diagnosing the basic properties and processes in those plasmas \citep{Brower02,Ding03}, observation of the nonlinear interaction of Alfv\'{e}n waves \citep{Carter06}, and a number of experimental efforts to investigate the nature of magnetic field reconnection, a core process in astrophysics \citep[e.g.][]{Brown02, Zweibel09,Yamada10}. The unity of plasma physics and plasma astrophysics is exemplified by the interesting fact  that the same radio propagation techniques, with the same radio telescopes, are, or can be, used to study the plasma physics of the interstellar medium, the corona, and the solar wind.

\subsection{The Fundamentals: (1) Phases of the ISM} As has been noted for decades, the interstellar medium exists in a number of ``phases'' of different temperature, density, and ionization state.  These different properties mean that fundamental plasma parameters such as the ion gyroradius, ion cyclotron frequency, plasma $\beta$, and Debye length differ from one phase to another.  The different phases and their plasma parameters were discussed in \cite{Spangler01}. Since this paper will discuss a number of parts of the ISM, we list in Table 1 the phases of the interstellar medium, together with their physical properties. Column 1 gives the astronomical name for the phase, column 2 gives the number density, and column 3 the temperature.  Column 4 gives the plasma $\beta$ (discussion below), and column 5 gives the volume filling factor of each phase.  The numbers in this last column are taken directly from Table 1.1 of \cite{Tielens05}, with the exception of the value for the Very Local Interstellar Medium (VLISM), which is taken from \cite{Frisch11}. 

The physical parameters listed in Table 1 represent averages over large volumes, and in some cases are quite uncertain. The main point of this table is to illustrate the great variety of physical conditions in the ISM.  One of the best diagnosed phases in Table 1 is that of the VLISM, consisting of a group of clouds within about 15 parsecs of the Sun.  Their properties are known well from high resolution spectroscopy as well as studies of the interaction of the heliosphere with these clouds.  A discussion of the properties of the VLISM, as well as the means for deducing these characteristics, is given in \cite{Frisch11} and \cite{Redfield09}. 

The plasma $\beta$ is an important parameter in specifying the nature of any plasma.  It is usually defined as \citep{Krall73}
\begin{equation}
\beta \equiv \frac{p}{B^2/8 \pi}
\end{equation} 
where $p$ is the gas pressure and $B$ is the magnitude of the magnetic field.  An alternative, and sometimes more meaningful definition is in terms of two fundamental wave speeds in a plasma, the ion acoustic speed $c_s$ and the Alfv\'{e}n speed $V_A$ \citep{Spangler97}
\begin{equation}
\beta \equiv \frac{c_s^2}{V_A^2}
\end{equation} 
where $c_s = \sqrt{\frac{\gamma k_B (T_e + T_i)}{m_i}}$ \citep{Nicholson83} and $V_A = \frac{B}{\sqrt{4 \pi \rho}}$.  In the above definitions, $\gamma$ is the ratio of specific heats (taken to be 5/3 for the calculations below), $k_B$ is Boltzmann's constant, $T_e$ and $T_i$ are the electron and ion temperatures, respectively, $m_i$ is the mass of the ion which constitutes the gas (taken here to be hydrogen), and $\rho$ is the mass density in the plasma.  The two definitions of $\beta$ in Equations (1) and (2) are nearly identical, differing only by a factor of $\gamma$, so they are the same for an isothermal equation of state.  

The reason for defining $\beta$ in terms of wave speeds rather than pressures is that the definition of Equation (2) better suits the problem at hand, which is an understanding of turbulence in the plasmas of the interstellar medium.  The ratio of wave speeds in Equation (2) is critical in determining wave damping and instability properties, as well as other wave characteristics such as compressibility.  To the extent that turbulence in astrophysical medium may be modeled in terms of wave properties, this definition of $\beta$ is more appropriate.  

We employ a number of simplifications in calculating $c_s$, $V_A$, and $\beta$.  We assume $T_e = T_i$. Although this is not the case in the solar corona and solar wind, it is known to be the case for the Warm Ionized Medium \citep[WIM, ][]{Haffner09}, and is probably the case in the clouds of the VLISM \citep{Spangler11}. We also assume a pure hydrogen plasma, with the important exception of the molecular clouds (see below).  This choice avoids the sometimes complicated question of the degree of ionization of helium.  Finally, a value of $B = 4 \times 10^{-6}$ G is chosen, with the exception of the molecular clouds.  

An important restriction in our calculations is that $\beta$ is calculated for the ``ionized fluid'', i.e. the gas that consists of electrons and ions.  This restriction is not meaningful for fully-ionized media like the solar corona and the WIM, but is an important point for partially-ionized media like the VLISM and molecular clouds.  This distinction most directly affects the Alfv\'{e}n speed via the choice of $\rho$. In Table 1 we choose $\rho$ to be the mass density of the ionized fluid, not the total density that includes the neutral gas.  Our choice is justified for plasma waves and fluctuations with size scales much smaller than the ion-neutral collisional scale.  For much larger scales corresponding to outer scales of turbulence in partially-ionized plasmas, neutral gas participates in the dynamics of the ionized fluid.  The effective Alfv\'{e}n speed is then lower, and the plasma $\beta$ higher.  

We have omitted values of $\beta$ for the Warm Neutral Medium (WNM) and Cold Neutral Medium (CNM).  At an excessively superficial level, one might think that the plasma $\beta$ is not a meaningful parameter for these neutral gases.  In reality, these phases will be ionized at some level.  In fact, an interesting recent contribution to the discussion of phases of the ISM has been the advocacy of Heiles for a ``fifth phase'', which is partially ionized \citep{Heiles11}. It seems likely that the ``fifth phase'' of Heiles is the same as the WNM, with perhaps an elevated degree of ionization due to the proximity to an HII region.  However, at the present time, the nature and characteristics of this partially ionized medium are not sufficiently specified to add to Table 1.  

Given the comments in the previous paragraph, it might seem odd to include a full set of entries for the molecular cloud phase, which contains cold, predominantly neutral, and molecular as opposed to atomic gas.  At the outset, it must be recognized that there is an enormous range of gas properties within the category of molecular clouds, from diffuse molecular gas, to dense cores, to protostars.  All derived parameters such as $c_s$, $V_A$, $\beta$, and filling factor also have an enormous range.  For this reason, we have chosen one restricted but well-discussed case in considering the plasma $\beta$.  

At a very basic level, it is obvious that molecular cloud gas is partially ionized because one of the most important observational diagnostics is line radiation from molecular ions such as HCO$^+$, H$_3^+$, and N$_2$H$^+$.  In an insufficiently appreciated paper, \cite{Smith92} uses data from millimeter wavelength observations of dense clouds to determine properties such as electron density and temperature, and then proceeds to show that these properties satisfy the classic criteria for the plasma state, such as a large number of electrons per deBye sphere.  In Table 1, our value of $\beta$ for molecular clouds is calculated for the parameters in \cite{Smith92}.  

An important result from Table 1 is the wide range of $\beta$ in the different phases of the ISM.  It is one of the reasons why the nature of turbulence in these different media may differ as well.  The extremely low value in molecular clouds warrants immediate comment.  The value for $\beta$ is so low because, relative to other ISM plasmas, molecular clouds have very low temperatures, high magnetic field strength \citep[for those clouds with H$_2$ densities greater than about 200 cm$^{-3}$ ][]{Crutcher10}, and low ion densities due to low ionization fractions.  The value for $\beta$ here suggests that small scale turbulent fluctuations or plasma waves that do exist in the ionized fluid of molecular clouds will have the properties of waves in low or zero $\beta$ plasmas.  However, for turbulent fluctuations and waves on much larger scales that involve motion of the neutral fluid as well, both the Alfv\'{e}n speed and $\beta$ will be much lower.     
  
\begin{table}
\caption{Main Phases of the Interstellar Medium}
\begin{tabular}{lllll}
\hline\noalign{\smallskip}
Astronomical Name & Density (cm$^{-3}$) & Temperature (K) & $\beta$ & filling factor  \\
\noalign{\smallskip}\hline\noalign{\smallskip}
Molecular Cloud & $200 - \geq 10^5$ & $\leq 100$ & $9 \times 10^{-8}$ & 0.050 \%  \\
Cold Neutral Medium (CNM)& 10 - 100 & $\sim 100$ & \ldots & 1 \% \\
HII regions & 5 - 10 & 8000 & 15 - 30 & $3 \times 10^{-3}$ \% \\
Warm Neutral Medium (WNM) & 0.1 - 0.5 & 8000 & \ldots & 30 \% \\
Warm Ionized Medium (WIM)& 0.1 - 0.5 & 8000 & 0.29 & 25 \% \\
Very Local Medium (VLISM) & 0.11 & 6700 & 0.27 & 6 - 19 \% \\
Coronal (HIM) & $5 \times 10^{-3}$ & $10^6$ & 1.8 & 50 \% \\
\noalign{\smallskip}\hline
\end{tabular}
\end{table}

Of the media listed in Table 1, the Warm Ionized Medium (WIM) is perhaps the one of greatest interest in plasma astrophysics.  This situation is due to the substantial body of observational data on this medium; it is probably the best-diagnosed astrophysical plasma beyond the solar wind.  A major contribution to our understanding of the properties of the WIM has been the long term program of observing the medium with imaging Fabry-Perot interferometers operating in the H$\alpha$ line and other important spectral lines. This program was conceived and directed by R.J Reynolds of the University of Wisconsin; as a consequence the WIM is often referred to as the ``Reynolds Layer''. In the past decade, these  H$\alpha$ observations have been greatly expanded by the Wisconsin H-Alpha Mapper (WHAM) instrument, under the direction of L.M. Haffner.  An excellent review of the scientific results emergent from WHAM and a relevant bibliography is given in \cite{Haffner09}.  Observations complementary to those of WHAM are provided by radio propagation measurements, primarily pulsar observations but also in some cases of extragalactic radio sources.  Assembly of data on radio wave scattering of pulsars and extragalactic radio sources has led to the inference of the power spectrum of density fluctuations in the WIM \citep{Armstrong95}. This topic is discussed further in Section 1.2 below.
 
As has long been noted, the pressures of the less dense phases of the ISM are, very roughly, comparable at a value of $1.0 \times 10^{-13} - 1.0 \times 10^{-12}$ dynes/cm$^2$ \citep{Ferriere98, Tielens05}. By this standard, molecular clouds and HII regions are overpressured.  In the case of molecular clouds, the gravitational potential contributes to confinement of the gas.  HII regions are overpressured, expanding entities.  The other phases, Cold Neutral Medium (CNM), Warm Neutral Medium (WNM), Warm Ionized Medium (WIM), Very Local Interstellar Medium (VLISM) and Coronal Phase or Hot Ionized Medium (HIM) have roughly comparable pressures and may, in fact, be in pressure equilibrium.  In addition, the magnetic pressure of the interstellar medium is comparable to the aforementioned gas pressures, with a value of $\simeq 6.4 \times 10^{-13}$ dynes/cm$^2$ for $B_{ISM} = 4 \times 10^{-6}$ G.  Finally, the pressure corresponding to the energy density of the Galactic cosmic rays is also similar, $\simeq 1.0 \times 10^{-12}$ dynes/cm$^2$, suggesting equilibration between the forms in which the ISM can store energy.  This whole situation is summarized in the textbook by \cite{Tielens05}, where a value of  $\simeq 0.5 \times 10^{-12}$ dynes/cm$^2$ is quoted for the gas phases of the ISM (CNM, WNM, WIM, and HIM), and  a pressure of $\simeq 1.0 \times 10^{-12}$ dynes/cm$^2$ is assigned to both the magnetic and cosmic ray pressure. 

The pressures and other properties of the various phases of the ISM, as well as the pressures of the interstellar magnetic field and cosmic rays  were considered in detail by \cite{Ferriere98}.  \cite{Ferriere98} also estimated how these pressures change with Galactocentric radius and altitude above the Galactic plane.  For the Galactic location of the Sun, and in the Galactic plane, \cite{Ferriere98} estimates (see Figure 3 of that paper) a pressure of  $6.0 \times 10^{-13}$ dynes/cm$^2$ for the gas phase, and  $\simeq 1.0 \times 10^{-12}$ dynes/cm$^2$ for both the magnetic and cosmic ray pressures.  

The rough similarity between thermal gas pressure and magnetic pressure that seems to characterize the local Galactic ISM might not be universal.  \cite{Beck07} discusses observations and analysis of the galaxy NGC 6946, characterized by a high star formation rate, and reports results on the variation of all pressures with galactocentric distance.  \cite{Beck07} finds that the magnetic pressure is considerably larger than the value quoted above for the Milky Way, and that the ionized gas pressure is approximately an order of magnitude less than magnetic pressure.  The interstellar plasma of NGC 6946 appears to be a low-$\beta$ plasma, and the processes of energy equilibration have not progressed to completion.  

The above considerations are relevant to the scope of this paper.  Our view of the interstellar medium considers it as a dynamic plasma.  Magnetohydrodynamics (MHD) describes the dynamics as an interaction of the gas and the magnetic field via pressure terms as discussed above and magnetic tension forces.  In addition, plasmas interact with energetic particles such as the cosmic rays through resonant interactions with plasma turbulence.

Among the important, recent developments in this field has been continued progress in specifying the strength of the interstellar magnetic field $\vec{B}_{ISM}$ , and its dependence on gas density \citep{Crutcher10}. A plot of magnetic field strength (largely deduced through Zeeman effect measurements) versus gas density shows considerable scatter, and a trend towards larger values only for densities greater than about 200 cm$^{-3}$.  \cite{Crutcher10} also infer a median magnitude of the interstellar magnetic field in the low density phases of the ISM of $6 \pm 1~\mu$G, slightly higher than the value used in the calculations above. This is in agreement with equipartition estimates of the total magnetic field strength from synchrotron emission, and a factor of about three higher than the regular magnetic field component. At the present, there is no observational evidence for a change in the magnitude of $\vec{B}_{ISM}$ between different phases of the low density ISM.

\subsection{ The Fundamentals: (2) Radio Wave Propagation through the ISM} 

This paper will concentrate on two radio propagation effects, acting on small ($10^2 - 10^4$ km) scales and large (pc) scales, respectively. These are angular broadening of a compact or pointlike source due to density turbulence in the interstellar medium, and Faraday rotation of linearly polarized radio waves from a radio source embedded in the ISM, or outside the Galaxy. We also briefly allude to other radio scintillation phenomena caused by small scale turbulence. In the latter topic, we include the signature of ISM turbulence in gradients of the polarization vector of synchrotron radiation.

\subsubsection{Angular broadening of compact sources} 

The basic physics content of radio wave propagation through the ISM is to be found in the expression for the refractive index of radio waves in a plasma. This is discussed in the proceedings of previous meetings of the International Space Science Institute, i.e. \cite{Spangler01} and \cite{Spangler09}.  As discussed there, the refractive index depends on the plasma density and magnetic field. For radio wave propagation in the ISM, the magnetic field dependence is determined by the component of the magnetic field in the direction of wave propagation. The modification of the radio refractive index by the magnetic field is much smaller than the modification due to the plasma density. This is responsible for the well-known feature that radio propagation measurements primarily diagnose the plasma density of the ISM, with only a higher order contribution due to the Galactic magnetic field.

\begin{figure}
\centerline{\includegraphics[width=22pc]{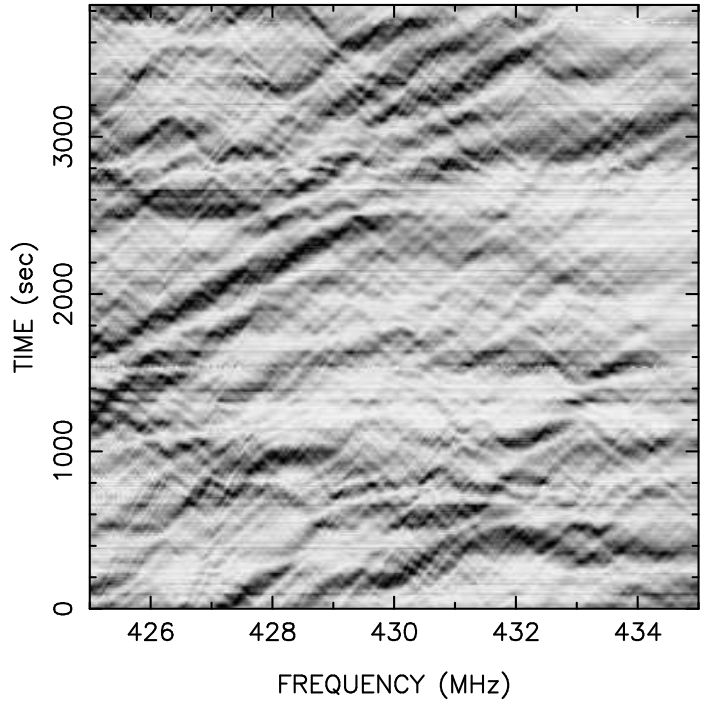}}
\caption{A dynamic spectrum of the pulsar PSR1133+16.  Wave propagation through the stochastic interstellar medium produces constructive and destructive interference at different spatial locations (here portrayed as time) and frequencies.  Observations and figure provided by James Cordes, Cornell University.  }
\label{fig:dynamic_spectrum}
\end{figure}

Turbulent fluctuations in the plasma density and magnetic field cause stochastic spatial and temporal fluctuations in the refractive index in the ISM.  As a result, propagation through such a medium induces all manner of fluctuations in the received radio wave field \citep{Uscinski77}. The theory of how fluctuations in the refractive index generate corresponding fluctuations in properties of the wave electric field (various n-point correlations of the electric field) is generally attributed to \cite{Tatarski61}. An excellent illustration of the effects of wave propagation through a random medium is given by dynamic spectra of pulsars, an example of which is shown in Figure~\ref{fig:dynamic_spectrum}\footnote{We thank James Cordes of Cornell University for providing this graph.} The spectrum of a pulsar is a measured as a function of time, and the set of spectra combined as shown in Figure~\ref{fig:dynamic_spectrum}.  In the absence of the turbulent interstellar medium, the flux density of the pulsar would be constant over the frequency range shown.  The gray scale indicates the brightness of the pulsar, with dark shaded regions being bright.  The variation in brightness is due to scattered radio waves alternatively constructively and destructively interfering at different frequencies and times. A discussion of pulsar dynamic spectra and the information they contain is given in \cite{Cordes86}.  The specific observations shown in Figure 1 are discussed in \cite{Lazio04}.  
 
A major goal of the theory of wave propagation in a random medium is to relate, via an integral transform, the radio astronomical measurement as a function of the independent variable, to the density power spectral density as a function of wavenumber.
 Examples of major contributions to this literature are \cite{Uscinski77}, \cite{Lee75}, and \cite{Rickett77,Rickett90} .  An illustration of the various types of observable stochastic propagation phenomena is given in Figure~1 of \cite{Spangler09}.

\subsubsection{Depolarization and Faraday Rotation of Synchrotron radiation}

Interstellar radio propagation measurements also provide information on the basic plasma state of the ISM, such as the plasma density, the vector magnetic field, as well as how these fields vary with position in the Galaxy.   

Since variations in the magnetic field vector along the line of sight and/or within the angular size of a telescope beam will partially depolarize linearly polarized synchrotron emission, the observed degree of polarization traces the ratio of large-scale (regular) magnetic field strength to total magnetic field strength \citep{Beck01}. However, due to small-scale variations in this ratio caused by local structure (supernova remnants, variable turbulence parameters), this method is mostly utilized on kpc-scales in external galaxies. 

Parsec-size scales in the magnetized ISM are typically probed using Faraday rotation. The Faraday rotation measure $RM$ is directly proportional to the path integral along the line of sight (los) of the electron density $n_e$ and line-of-sight component of the magnetic field $B_{\parallel}$:

\begin{equation}
  \left(\frac{RM}{\mbox{rad m}^{-2}}\right) = 0.81 \int_{los} \left(\frac{n_e}{\mbox{cm}^{-3}}\right) \left(\frac{B_{\parallel}}{\mu\mbox{G}}\right) \left(\frac{dl}{\mbox{pc}}\right)
\label{e:definition_rm}
\end{equation}
Measurements of $RM$ therefore provide nearly unique information on the magnetic field in the tenuous, ionized component of the ISM. Equation (3) illustrates the fact that the measured $RM$ is sensitive to the distribution of $n_e$ and $B_{\parallel}$ along the line of sight.  If $\vec{B}$ rotates through a large angle such that $B_{\parallel}$ changes substantially or even reverses sign along the line of sight, the value of $B$ inferred by the $RM$ is much less than the true value that would be measured in-situ.  This will also be the case if $n_e$ and $B_{\parallel}$ are anticorrelated in the medium being probed, as discussed by \cite{Beck03}.    

\begin{figure} 
\centerline{\includegraphics[width=22pc]{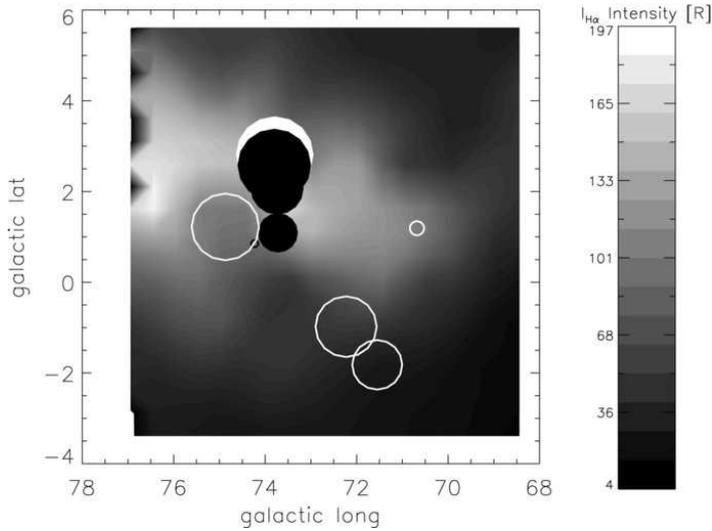}}
\caption{Highly variable Faraday Rotation Measures in the Galactic plane in Cygnus, in the vicinity of the Cygnus OB1 star formation region. The sizes of the plotted circles are proportional to the magnitude of $RM$, specifically proportional to the logarithm of the absolute magnitude of $RM$.  Filled circles correspond to positive $RM$, open circles are negative $RM$. The largest value of (absolute magnitude) of $RM$ is 732 rad/m$^2$, and the smallest is 7.5 rad/m$^2$. The gray scale represents the intensity of $H\alpha$ emission in this region; the $H\alpha$ data may show the upper half of the plasma shell associated with the Cygnus OB1 association.  Further discussion of these data is given in \cite{Whiting09}.  Figure taken from \cite{Whiting09}, reproduced by permission of the American Astronomical Society.}
\label{fig:rm_whiting}
\end{figure}

Traditionally, Faraday rotation is measured from the rotation of linear polarization angle $\theta$ as a function of wavelength $\theta = \theta_0 + RM \lambda^2$, where $\theta_0$ is the intrinsic polarization angle at emission of the synchrotron radiation.  An illustration of these ideas is given in Figure~\ref{fig:rm_whiting}, which shows the Faraday rotation measure $RM$ along several lines of sight to extragalactic radio sources in the Galactic plane in Cygnus. The Faraday rotation of the synchrotron radiation emitted by these sources is dominated by the Milky Way. The large differences in magnitude, and even sign, of $RM$ between closely-spaced lines of sight in Figure~\ref{fig:rm_whiting} are an indicator of the role of young, luminous stars in this region, as they produce structure in the ISM via stellar winds and supernovae, and ionize the gas.  

However, if synchrotron emission and the Faraday-rotating medium are mixed or alternating along the line of sight, the simple linearity of polarization angle change with $\lambda^2$ is no longer valid. This may be the case in the majority of Faraday rotation measurements. Faraday rotation measurements of the diffuse synchrotron emission in galaxies is the most obvious example, but extragalactic sources may also have several intrinsic RM components. In this case, every RM component $i$ along the line of sight - now called Faraday depth $\phi_i$ to indicate that it only probes Faraday rotation along a part of the line of sight - adds its own polarization angle rotation of $\phi_i\lambda^2$, resulting in a non-linear polarization angle change with $\lambda^2$. However, this opens the possibility of a Fourier transform, with $\lambda^2$ as one of the conjugate variables, in order to disentangle the various $\phi_i$ components in the total observed signal. This method is called Rotation Measure synthesis \citep{Burn66,Brentjens05}. Rotation Measure synthesis takes as its basic observable field a complex polarization function formed from the Stokes parameters $Q$ and $U$, $P \equiv Q + \imath U$.  The observable $P$ is a function of wavelength, or wavelength squared.    

This Fourier transform relation between the observed polarization function as a function of wavelength squared $P(\lambda^2)$ and the Faraday dispersion function (or Faraday spectrum) is expressed as

\begin{eqnarray}
P(\lambda^2) &=& \int_{-\infty}^{+\infty} F(\phi) \mbox{e}^{2i\phi\lambda^2} d\phi \\
F(\phi) &=& \int_{\infty}^{+\infty} P(\lambda^2)\mbox{e}^{-2i\phi\lambda^2} d\lambda^2
\end{eqnarray}

However, since integration over wavelengths from $-\infty$ to $+\infty$ is not possible by definition, in practice these equations include a window function $W(\lambda^2)$ which is non-zero where there is wavelength coverage and zero elsewhere:

\begin{eqnarray}
P_{obs}(\lambda^2) &=& W(\lambda^2)P(\lambda^2) = W(\lambda^2)\int_{-\infty}^{+\infty} F(\phi) \mbox{e}^{2i\phi\lambda^2} d\phi \\
F_{obs}(\phi) &=& F(\phi)\ast R(\phi) = K\int_{\infty}^{+\infty} P_{obs}(\lambda^2)\mbox{e}^{-2i\phi\lambda^2} d\lambda^2 \,\,\,\mbox{where} \\
K &=& \left(\int_{\infty}^{+\infty} W(\lambda^2) d\lambda^2\right)^{-1}
\end{eqnarray}

This introduces the Rotation Measure Spread Function (RMSF) $R(\phi)$, which describes sidelobes in the Faraday depth signal due to imperfect wavelength coverage, very analogous to the dirty beam (or point spread function) in radio interferometry due to imperfect coverage of the $(u,v)$ plane.

\begin{figure}
\centerline{\includegraphics[width=0.8\textwidth]{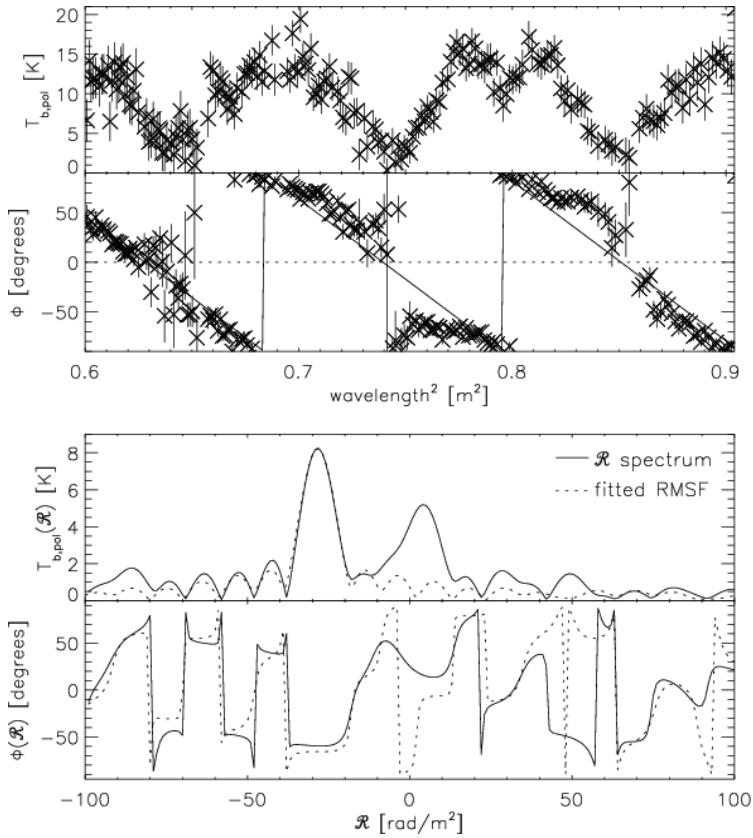}}
\caption{Polarized intensity (top) and polarization angle (2nd from top) as a function of wavelength squared for synchrotron emission from an extragalactic radio source observed by \cite{Schnitzeler09}. The Faraday spectrum (3rd from top) as a function of Faraday depth $\phi$ shows two peaks, indicating synchrotron-emitting source components at two separate Faraday depths along this line of sight. Credit: Schnitzeler et al, A \& A 494, 611, 2009, reproduced with permission \copyright ESO.}
\label{fig:rms_example}
\end{figure}

A nice illustration of this effect is given in Figure~\ref{fig:rms_example}, which is taken from \citet{Schnitzeler09}. The Figure shows polarized synchrotron intensity and polarization angle as a function of wavelength squared of an extragalactic radio source. Variations in polarized intensity and non-linearity of polarization angle with $\lambda^2$ suggest that synchrotron emission and Faraday rotation are partially mixed. Indeed, the Faraday spectrum in the Figure shows two Faraday depth peaks belonging to two different synchrotron emitting regions with different amounts of  Faraday rotation along the line of sight. The two Faraday depth components might be due to a magneto-ionized medium in or around the extragalactic point source. Alternatively, both Faraday depth components could be due to the Galactic interstellar medium, with one of the emission components also being Galactic.  In the latter case, the polarized flux of the Galactic component would have to fortuitously be comparable to the extragalactic source. Without additional information it is not possible to distinguish between these two possibilities.

\section{Different Astrophysical Media, Common Phenomena}  

It is a remarkable fact that a number of astrophysical plasmas, i.e. distinct and very different media, have roughly similar radio propagation effects. The Faraday rotation measure ($RM$) due to the Earth's ionosphere is typically in the range $0.5 - 3 \mbox{ rad/m}^2$.  This is of the order of the $RM$ due to the solar corona for lines of sight that pass at a heliocentric distance $\sim 10~R_{\odot}$, and a factor of a few smaller than the standard deviation in $RM$ of the interstellar medium at high Galactic latitudes \citet[e.g.][]{Mao10}.  

Another example is provided by Very Long Baseline Interferometry.  Very Long Baseline Interferometers operating at frequencies of 1 - 5 GHz measure similar effects, and of similar magnitude, when observing extragalactic radio sources through the inner solar wind at heliocentric distances of 10 - 30~$R_{\odot}$ or through the Galactic plane \citep{Spangler95,Spangler02,Spangler88,Fey91,Spangler98}.

The fact that similar observational effects are measured for very different media means that scientists who study the ISM should be in dialog with heliospheric scientists who are studying related scientific questions.  Finally, in the case of Faraday rotation, there is also a connection with laboratory plasma research.  Faraday rotation is used as a diagnostic of fusion plasmas \citep{Brower02,Ding03}.  This offers ISM astronomers the possibility of laboratory ``ground truth'' for some of our diagnostics.

\section{Well-Established Results from Radio Propagation Studies}
There are several results from radio propagation studies that are so well established, and confirmed by different investigators, that they now have the status of basic properties of the ISM that should be explained by theories.  
\subsection{``The Big Power Law in the Sky''}
One of the best-known results from radio propagation studies is that the spatial power spectrum of density fluctuations follows a Kolmogorov spectrum over at least 5, and perhaps 10 decades \citep{Armstrong95}. It should be emphasized that the result of \cite{Armstrong95} pertains to the Warm Ionized Medium (WIM) component of the ISM.  
Analogous results have been reported for diagnostics of the neutral gas \citep[e.g. ][]{Chepurnov10b}.  The neutral gas resides in the WNM phase, which is spatially distinct from the WIM. As noted above, (Section 1.1) the degree of ionization in the WNM, while low, is not zero, and it may be considered as a weakly-ionized plasma. As such, the dynamics of the plasma component are relevant, since they are communicated to the neutrals via collisions.  Whether the fluctuations studied by \cite{Chepurnov10b} are driven by plasma dynamics due to the minority ion species remains unknown.  

It is intriguing that the solar wind density power spectrum is the same as that observed for the ISM (at least for the slow solar wind), although the inertial subrange is smaller, being perhaps 3 orders of magnitude \citep{Bruno05}.  In the case of the ISM, it is perhaps not sufficiently appreciated that there must be a corresponding power law spectrum of magnetic field fluctuations to explain the highly diffusive transport properties of cosmic rays with a very wide range of energies.  This point has been made and emphasized by \cite{Jokipii77,Jokipii88}.

\subsection{Spatial Variation in the Intensity of Interstellar Turbulence} A measure of the intensity of turbulence is the parameter $C_N^2$, which is the normalization constant of the density power spectrum.  That is, if the spatial power spectrum of density fluctuations is $P_n(q)$ where $q$ is the spatial wavenumber, 
\begin{equation}
P_n(q) = C_N^2 q^{-\alpha}
\label{e:density_power_spectrum}
\end{equation}
and $\alpha$ is the spectral index of the power spectrum. This form of the power spectrum is the simplest, in which the power spectral density depends only on the magnitude of $\vec{q}$, and is thus isotropic.  It is known that magnetohydrodynamic (MHD) turbulence is in fact anisotropic, with the large scale magnetic field determining the preferred direction (See comments in Section 3.4 below).  In this case, the power spectral density $P_n(\vec{q})$ depends separately on $q_{\perp}$ and $q_{\parallel}$, the components of $\vec{q}$ perpendicular and parallel, respectively, to the large scale field \citep[this is discussed in ][with a guide to the relevant literature]{Spangler99}. For most of this paper, we adopt Equation~\ref{e:density_power_spectrum} as a convenient approximation.

 The parameter $C_N^2$ is directly related to the variance of the density fluctuations.  It has been long realized that $C_N^2$ varies drastically from one part of the interstellar medium to another \citep{Rickett77,Cordes85}.  In some cases, it is clear that lines of sight with large $C_N^2$ traverse HII regions, or other regions with higher than normal plasma density.  However, there appear to other lines of sight where no such obvious region of enhanced density exists.  It remains unclear whether some of this variation could be due to true turbulent intermittency \citep{Spangler98,Spangler99}.

\subsection{The Galactic-Scale Magnetic Field} 

Large-scale Faraday rotation surveys of the sky show substantial organization of the $RM$ in different parts of the sky, in the sense that the magnitude and sign of $RM$ are correlated over significant parts of the sky. This is interpreted as evidence of a Galactic-scale magnetic field. Most likely, the large-scale magnetic field in the Milky Way generally follows the spiral arms, as is ubiquitously seen in synchrotron observations of external galaxies \citep{Beck01}. However, there is some evidence for local deviations from the spiral structure \citep{Brown07,Rae10, Vaneck11}, similar to M51 \citep{Patrikeev06}. One reversal in the large-scale magnetic field direction just inside the Solar circle has been known for decades \citep[e.g.][]{Thomson80}, although it remains unclear why these large-scale reversals are not observed in external spirals. There is still much controversy about the number and location of any other large-scale field reversals in the Galactic disk. For an extensive review, see \citet{Haverkorn13}.

\subsection{Anisotropy of Turbulence}

A major result from the theory of magnetohydrodynamic (MHD) turbulence, confirmed by observations of turbulence in the solar corona and solar wind, is that turbulence is anisotropic, in the sense that turbulent irregularities are stretched out along the large scale magnetic field. This result was obtained by \cite{Strauss76} for the case of irregularities in fusion plasmas, but the arguments presented by Strauss are also valid in the case of astrophysical plasmas. This has been broadly appreciated in the astrophysical, as well as plasma physics community since the work of \cite{Goldreich95}, which advocated a view of MHD turbulence as comprised of counterpropagating Alfv\'{e}n waves. The interaction of these Alfv\'{e}n waves consequently develops a perpendicular cascade of turbulent energy.  This anisotropy is also observed to be present in interstellar turbulence \citep[see][for recent discussions of the more pronounced cases]{Brisken10,Rickett11}.

\subsection{The Dissipation Range of Turbulence}
\label{s:dissipationrange}

Important progress has been made in the past decade in our understanding of the dissipation of plasma turbulence.  This progress has been possible through improved measurements of solar wind turbulence, as well as novel theoretical developments \citep{Howes08,Alexandrova09,Alexandrova12,Howes11,Howes11b,Sahraoui12}.  These investigations have identified spatial scales on which dissipation occurs, and advanced suggestions for the responsible mechanisms.  It is now clear that a break in the power spectrum of magnetic field fluctuations in the solar wind occurs on scales comparable to, and smaller than the ion inertial length $l_i$, 
\begin{equation}
l_i \equiv \frac{V_A}{\Omega_i}
\label{e:ion_inertial_length}
\end{equation}
where $V_A$ is the Alfv\'{e}n speed and $\Omega_i$ is the ion (proton) cyclotron frequency \citep{Howes11,Alexandrova09,Alexandrova12}. There remains active discussion in the community as to whether the dissipation is due to Landau damping of highly oblique Alfv\'{e}n waves (the assumption being that turbulent fluctuations on these scales have damping properties similar to linear plasma wave modes \citep{Howes08,Howes11,Howes11b}), dissipation of other, higher frequency modes propagating at large angles with respect to the mean magnetic fields \citep{Sahraoui12}, or damping by other modes on electron inertial scales \citep{Alexandrova09,Alexandrova12}.  In keeping with the philosophy of this paper, we assume that these results are of great importance of our understanding of the ISM as well. 

As will be discussed in Section 4, there is good evidence for the beginning of the dissipation range in interstellar turbulence at the ion inertial length, although there is also evidence for differences between solar wind and interstellar turbulence (see Section 4.5 below).

\subsection{The Outer Scale of Interstellar Turbulence}
The outer scale to magnetized turbulence in the Warm Ionized Medium (WIM) phase of the ISM is measured from structure functions of RM to be a few parsecs in the spiral arms, but up to $\sim100$~pc in the interarm regions \citep{Haverkorn06,Haverkorn08}. This is in agreement with estimates of the outer scale of turbulence averaged over large parts in the sky (mostly towards the Galactic halo) of order 100~pc by \citep[e.g.][]{Lazaryan90, Ohno93,Chepurnov10}. These scales are similar to the final sizes of supernova remnants, suggesting (combined with energy arguments, see \citet{MacLow04}) that these are the dominant sources of turbulence in the WIM. The smaller outer scale found in spiral arms may be due to the abundance of H~{\sc ii} regions \citep{Minter96,Haverkorn04}. Small outer scales of a few parsecs have also been found in the highly polarized Fan region \citep{Iacobelli13} or from anisotropies in TeV cosmic ray distributions \citep{Malkov10}. 

\section{Turbulent Microscales in the Interstellar Medium}
\subsection{Definition of Turbulent Microscales}
In this section, we discuss the ways we can measure turbulence on very small scales in the interstellar medium.  By very small scales, we mean those in the dissipation range.  Knowledge of this part of the turbulent cascade is very important because it contains information on the way in which energy is taken from the large scales and transferred to other forms, presumably heat energy of the interstellar gas.  

Our interest in this section will be particularly focused on two phases of the ISM, the WIM and HII regions surrounding young stars.  The WIM is of interest because it appears to be the best-diagnosed phase of the ISM, as discussed in Section 1.1 above. 
\subsection{Turbulent Microscales in the Solar Wind} Once again, we use the solar wind, with its extensive and often sophisticated, in-situ measurements and substantial body of theoretical results as a model for interstellar turbulence. Spacecraft instruments provide in-situ measurements of virtually all plasma parameters of interest, and provide the best data set for discussions of MHD turbulence.  Measurements of solar wind turbulence at a heliocentric distance of 1 a.u. (the bulk of spacecraft measurements) show a power-law power spectrum of magnetic field fluctuations with a single spectral index that extends from an outer scale with a size of one to a few solar radii, down to an inner scale of a few thousand kilometers \citep[e.g.][]{Bruno05}. A number of investigations have shown that this scale corresponds to the ion inertial length $l_i$ defined in Equation~\ref{e:ion_inertial_length}.

Radio propagation observations through the corona and inner solar wind \citep{Coles89,Harmon05,Spangler95} also show strong evidence for an enhancement in the power spectral density of plasma density fluctuations on the scale of the ion inertial length.  This bulge on the approximate scale of the ion inertial length can also be seen in power spectra from in-situ measurements of plasma density in the solar wind at 1 a.u. \citep{Chen12}.
These observations, from direct, in-situ measurements as well as radio propagation observations,  are interpreted as evidence that the fluctuations in the dissipation range have properties of obliquely-propagating, kinetic Alfv\'{e}n waves, since such waves become more compressive at the ion inertial scale (Harmon 1989).  In fact, the kinetic Alfv\'{e}n nature of the fluctuations on the dissipation scale is the basis of the model of turbulence in the dissipation range advanced by Howes and colleagues \citep{Howes11}.  The situation for the solar corona and solar wind seems quite consistent as regards both measurements of plasma density fluctuations and theoretical understanding of the entire turbulent cascade. The prominence of this directly-detected bulge appears to be less pronounced than that retrieved from radio propagation measurements in the corona and inner solar wind.  This may indicate that the kinetic Alfv\'{e}n wave component decays with increasing heliocentric distance in the solar wind.  This would hardly be surprising, since many properties of the solar wind change with heliocentric distance \citep[ e.g.][]{Bruno05}. 

\subsection{How We Measure Microscales in Interstellar Turbulence} 
The ion inertial length in the WIM phase of the interstellar medium is of order one hundred to a few hundred kilometers \citep{Spangler90}.  At first, it seems amazing that any kind of astronomical measurement, made on a medium with an extent of kiloparsecs, could diagnose fluctuations on such a scale. Radio propagation measurements make this possible.

The subsequent discussion in this section will concentrate on one of the scintillation phenomena mentioned in Section~1.2, angular broadening.  A point source of radio waves viewed through a turbulent medium will appear as a blurred, fuzzy object.  Essentially the same phenomenon is encountered at optical wavelengths in the form of  seeing disks of stars.  A radio source that is blurred by interstellar turbulence will have a measured brightness distribution $I(x,y)$ which is more extended than the intrinsic image of the source. Here $I$ is the intensity of the radiation, which is a function of two angular coordinates on the sky, $x$ and $y$, customarily Right Ascension and Declination. This brightness distribution contains information on the intensity and spatial power spectrum of the density fluctuations.  The brightness distribution is related to the observable quantity which is directly measured by the interferometer. The way a radio interferometer makes an image of a radio source is to measure the {\em complex visibility function} $V(u,v)$, which is directly related to the correlation between the radio wave electric field at two antennas of an interferometer \citep{Thompson86}.  The complex visibility function has units of Janskys (radiative flux), and is a function of the arguments $u$ and $v$, the east-west and north-south components of the interferometer baseline, normalized by the wavelength of observation.  
A two dimensional Fourier transform relates the complex visibility function $V(u,v)$ and $I(x,y)$ \citep{Thompson86}.

Observations have shown that the turbulent irregularities in the WIM and HII regions around OB associations, like those in the corona and solar wind, are anisotropic in the sense that is theoretically expected. A summary of the observational evidence as of 1999 is given in \cite{Spangler99}. For the present purposes, we will employ the simplifying approximation of isotropic irregularities. The arguments given here can be generalized to the case of anisotropic scattering \citep{Spangler88}.
In the case of isotropic scattering, 
\begin{equation}
  V(u,v) = V(\sqrt{u^2 + v^2}) = V(r)
\end{equation}
where $r$ is the (dimensional) interferometer baseline length, projected on the plane of the sky. In this case, the complex visibility function of a point source viewed through a turbulent medium is a real function, \citep{Cordes85,Spangler90}
\begin{equation}
  V(r) = S_0 e^{-\frac{1}{2}D_{\phi}(r)}
  \label{e:vis_phase_sf}
\end{equation} 
where $S_0$ is the total flux density of the source, and $D_{\phi}(r)$ is the {\em phase structure function}, which contains information on the intensity and spatial power spectrum of the density fluctuations. For the case of scattering by a homogeneous, turbulent slab of thickness $L$, the phase structure function is \citep{Cordes85,Spangler90}
\begin{equation}
  D_{\phi}(r) = 8 \pi^2 r_e^2 \lambda^2 L \int_0^{\infty} dq q \left[ 1 - J_0(qr) \right] P_n(q) 
  \label{e:phase_sf_1}
\end{equation}
The functions and variables in Equation~\ref{e:phase_sf_1} are as follows.  The classical electron radius is given by $r_e$, $\lambda$ is the wavelength of observation, $L$ is the thickness, or extent along the line of sight of the turbulent plasma, $q$ is the magnitude of the turbulent wavenumber, $J_0(x)$ is a Bessel function of the first kind of order 0, and $P_n(q)$ is the spatial power spectrum of the density fluctuations. One expects the power spectra of all plasma parameters to be modified at wavenumbers corresponding to the reciprocal of the ion inertial length, ion gyroradius, or similar plasma microscale on which dissipation begins to become important. \cite{Spangler90} adopted the following simple model in which Equation~\ref{e:density_power_spectrum} is modified by having the power spectrum truncated on wavenumbers larger than a dissipation wavenumber $q_0$, 
\begin{equation}
  P_n(q) = C_N^2 q^{-\alpha} e^{-q/q_0}
  \label{e:density_power_spectrum_trunc}
\end{equation}

Although Equation~\ref{e:density_power_spectrum_trunc} is highly simplified, and was adopted by \cite{Spangler90} for analytic convenience, the power spectrum of magnetic field fluctuations in the solar wind at 1~AU is truncated by an exponential function \citep{Alexandrova09,Alexandrova12}.  An important difference between the result of \cite{Alexandrova09,Alexandrova12} and the analysis presented below is that Alexandrova and coworkers found exponential truncation of the power spectrum on electron rather than ion scales (i.e.\ electron inertial length, or electron gyroradius).

Substitution of Equation~\ref{e:density_power_spectrum_trunc} into Equation~\ref{e:phase_sf_1}, and change of variables from $q \rightarrow y \equiv qr$ gives the following expression 
\begin{equation}
  D_{\phi}(r) = 8 \pi^2 r_e^2 \lambda^2 ( C_N^2 L) r^{\alpha-2} \int_0^{\infty} dy  \left[ 1 - J_0(y) \right] y^{-(\alpha-1)} e^{-y/Q} 
  \label{e:phase_sf_2}
\end{equation}
where $Q \equiv \frac{r}{l_d}$, with $l_d$ being the dissipation scale, $l_d \simeq \frac{1}{q_0}$. The quantity $C_N^2 L$ is termed the {\em scattering measure}, and roughly determines the magnitude of angular broadening.  Turbulent plasmas that have large $C_N^2$, large $L$, or both, will produce heavy angular broadening. 

Equation~\ref{e:phase_sf_2} yields important insight on remote sensing diagnosis of interstellar (and heliospheric) turbulence.  Let us start with the case $Q \rightarrow \infty$, which corresponds to an infinitely small dissipation scale.  In this case, the spectrum is power law for all wavenumbers larger than that corresponding to the outer scale. In this case, the integral in Equation~\ref{e:phase_sf_2} is a number which depends only on the index $\alpha$.  For the Kolmogorov spectrum ($\alpha = 11/3$) the value of the integral is 1.117, and the structure function $D_{\phi}(r) \propto r^{5/3}$.

\begin{figure} \includegraphics[width=22pc]{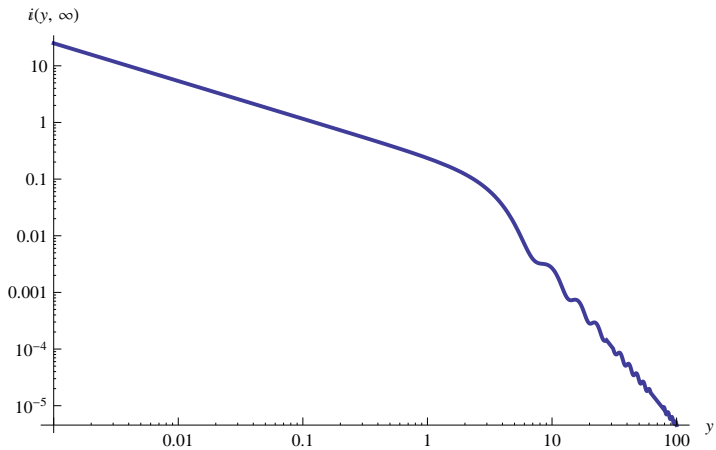} 
\caption{Plot of the function $I(y,Q)$ for a Kolmogorov spectrum ($\alpha = 11/3$) and $Q = \infty$.  The integrand tells what wavenumbers in the turbulent spectrum dominate the measurement on a given interferometer baseline.  This occurs for $y \equiv qr \geq 1$, which corresponds to irregularities with sizes of order the interferometer baseline.}
\label{fig:integrand_phaseSF}       
\end{figure}

Equation~\ref{e:phase_sf_2} also illustrates one of the most intriguing aspects of radio wave propagation, and demonstrates why radio astronomical measurements can contribute much to a discussion of plasma microscales in the ISM. Since Equation~\ref{e:phase_sf_2} is an integral over wavenumber (this is explicit in Equation~\ref{e:phase_sf_1}), the integrand shows which wavenumbers dominate the measurement.  The integrand in Equation~\ref{e:phase_sf_2}, which we note by the function $I(y,Q)$, is shown in Figure~\ref{fig:integrand_phaseSF} for the case $I(y,\infty)$ and $\alpha = 11/3$ (Kolmogorov spectrum).  

The function $I(y,\infty)$ is monotonically decreasing with increasing $y$.  At first, this would seem to indicate that the lowest wavenumbers in the spectrum dominate the measurement.  However, that is not true for the case of a Kolmogorov spectrum.  The measurement ($D_{\phi}$) is determined by an integral over $y$. This is because for $y \leq 2$, each progressively higher decade in $y$ makes a larger contribution to the integral. The integral is dominated by contributions with $y \sim 1 - 10$ where there is an inflection in the function $I(y,\infty)$. When an interferometer measures a broadened radio source, it is responding to irregularities with a wide range of wavenumbers.  However, the dominant contribution to the visibility measurement is from irregularities with sizes comparable to the baseline length.  This baseline length ranges from tens of kilometers in the case of the Very Large Array, to a few thousand kilometers in the case of Very Long Baseline Interferometers. Obviously, for this statement to be relevant, a measurement with a given interferometer must be affected or even dominated by propagation effects. This point was made in the context of interstellar scattering in \cite{Spangler88b}. 

Equation~\ref{e:phase_sf_2} also shows how the presence of turbulent dissipation is manifest in radio propagation measurements.  When $Q$ is finite, corresponding to a finite dissipation scale, the $e^{-y/Q}$ term will depress the value of the integral.  The value of $D_{\phi}(r)$ at short baselines, where dissipation is pronounced, is less than a value extrapolated from larger values of $r$ according to an $r^{5/3}$ relation.  In the dissipation range, $D_{\phi}(r)$ has a steeper dependence than $r^{5/3}$.  To illustrate these points, Figure 5 shows two structure functions, one without an inner scale and the other with an outer scale of 300 km.  

\subsection{Observational Results on Turbulent Dissipation Scales in the Interstellar Medium}

These issues were discussed in \cite{Spangler90}, who showed that observers who interpret their angular broadening data in terms of a spectral index $\alpha$ would report values which depend on the baselines used in the measurement.  Angular broadening measurements on short baselines would yield a value of $\alpha \simeq 4.0$, whereas measurements on long baselines in the inertial subrange would yield $\alpha \simeq 3.67$. This may be seen by reference to Figure 5.  The inferred spectral index is determined by the slope of $D_{\phi}(r)$ versus $r$ on a log-log plot, such as Figure 5.  \cite{Spangler90} assembled the data on angular broadening measurements that were available at that time, and showed that a dependence of the inferred value of $\alpha$ on the interferometer baselines used did seem to be present in the data.  The result was shown in Figure 1 of \cite{Spangler90}. From these data they found that there could be a break in the interstellar density power spectrum with an inner scale of 50 - 200 km.   More importantly, they pointed out that a scale in this range was actually expected, if the inner scale corresponds to the ion inertial length $l_i$ defined in Equation~\ref{e:ion_inertial_length}, as is the case for scattering in the corona and solar wind (Section 4.2).   

\begin{figure}
\includegraphics[width=22pc]{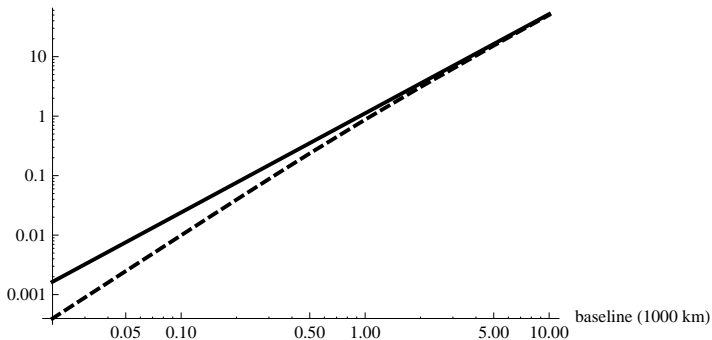}
\caption{Theoretical phase structure functions due to turbulence in the interstellar medium. The plotted lines correspond to theoretical structure functions $D_{\phi}(r)$ given by Equation (12).  Both adopt a Kolmogorov spectrum of density irregularities.  The solid curve is for a Kolmogorov spectrum with no inner scale ($l_d \rightarrow 0$). The dashed line corresponds to a spectrum with an inner scale $l_d = 300$ km. Sensitive, carefully calibrated interferometer measurements can distinguish between these two cases.}
\end{figure}

A more direct determination of an inner scale, made from comparison of measured $D_{\phi}(r)$ values with the theoretical expression in Equations (10) and (12), was made by \cite{Molnar95}. \cite{Molnar95} made and analyzed angular broadening measurements similar to those of \cite{Spangler88}, but of the radio source Cygnus X-1.  It is viewed through the HII region associated with the Cygnus OB2 association. \cite{Molnar95} essentially made a fit of Equations (12) and (15) (but also including the effect of anisotropy of scattering) to their data, and found a satisfactory model to be a Kolmogorov underlying spectrum and an inner scale of 300 kilometers.  

An additional, and particularly compelling result has been the recent report by \cite{Rickett09}.  They studied the form of the broadening profile of pulses from the pulsar PSRJ1644-4559. Late in the pulse, radiation is being received from highly scattered rays that are probing very small scale irregularities. \cite{Rickett09} make the important point that the amount of radiation received late in the pulse is only consistent with a Kolmogorov spectrum that breaks at an inner scale as expressed by Equation~\ref{e:density_power_spectrum_trunc}, or something similar.  A power spectrum that remained Kolmogorov to infinitely high spatial wavenumber would cause more pulse power to be observed late in the pulse than is actually seen. \cite{Rickett09} use their data to extract a value for the inner scale of 70 - 100 km.  

These three independent investigations using radio propagation data have therefore concluded that there is a spectral break in the density power spectrum of the interstellar medium, and that this inner scale is consistent with the ion inertial length. It should be emphasized that the lines of sight analyzed by \cite{Spangler90}, \cite{Molnar95}, and \cite{Rickett09} all traversed HII regions.  There is, as yet, no observational data that can determine the inner scale to the turbulence in the WIM. We do not know if such an inner scale would be at the ion inertial length. 

\subsection{A Break or a Bulge?}  

One interesting, and at the present preliminary feature emergent from these investigations regards the transition to the dissipation range in interstellar turbulence. In HII region plasmas, the dissipation range appears to consist of a smooth steepening, without the bulge in the density power spectrum on the ion inertial length, as exists in the corona and solar wind (Section 4.2). Given the admittedly limited present information, it appears that the interstellar spectrum of density fluctuations has no compressive bulge at the inner scale.  If confirmed by subsequent investigations, it could point to an important distinction between turbulence in the interstellar medium and that in the solar corona and solar wind. The results from \cite{Rickett09} seem particularly compelling, because a bulge in the interstellar density power spectrum on the ion inertial scale would produce more pulse power at late arrival times than is actually seen (this point is clearly illustrated in Figure~7 of \cite{Rickett09}).

If this bulge is missing in the interstellar density spectrum, what does it signify?  Does it imply that kinetic  Alfv\'{e}n waves are not present in the turbulent field, or that the small scale irregularities in the interstellar medium do not evolve in a manner similar to kinetic  Alfv\'{e}n waves?  In that case, what is the nature of the fluctuations over such a large inertial subrange in the ISM?  As mentioned in Section 4.2, the solar wind results may provide guidance; the results of \cite{Chen12} indicate that the prominence of kinetic Alfv\'{e}n waves decreases with increasing heliocentric distance.  Interstellar turbulence is comparatively much older in terms of the number of eddy turnover times, so it is certainly plausible that the kinetic Alfv\'{e}n wave component of ISM turbulence has dissipated. 

Another possible resolution is also suggested by studies of the solar wind.  As shown in Table 1, HII regions have large values of $\beta$, $\beta \gg 1$.  \cite{Chandran09}, in a discussion of solar wind density fluctuations, showed that the compressibility of kinetic  Alfv\'{e}n waves decreases with increasing $\beta$ \citep[see Figure 3 of ][]{Chandran09}. Kinetic Alfv\'{e}n waves may well be present in HII regions, but are relatively incompressive and make a small contribution to the density fluctuations in these plasmas.  

These ruminations need more extensive and more convincing observational demonstration.  Fortunately, the instruments and observational techniques are operational and available. The instruments currently available for angular broadening measurements are greatly improved over those used in the measurements cited above \citep{Spangler88,Spangler90,Molnar95,Spangler98}.  Those investigations used Very Long Baseline Interferometers with much smaller bandwidths and correlator capability than are now available with the Very Long Baseline Array (VLBA) of the National Radio Astronomy Observatory (NRAO). In addition, the LOFAR low frequency radio telescope in Europe is now operational and has the capability of making novel angular broadening measurements.  Finally, the work of \cite{Rickett09} also demonstrates the advances that have been made in pulsar measurements of the ISM, utilizing new, state-of-the-art pulsar processors on large single dish telescopes such as the Parkes antenna or the Green Bank Telescope of NRAO.  Future investigations with these powerful new instruments could illuminate the interesting question as to whether turbulent dissipation processes are the same in the solar wind and the plasma components of the interstellar medium. 

\section{Turbulent mesoscales in the interstellar medium}

\subsection{Rotation Measure Synthesis results}

The interpretation of three-dimensional Faraday depth cubes (i.e. polarization 
intensity maps in spatial coordinates where Faraday depth is the third
dimension) is anything but straightforward.

In addition to the artefacts introduced by a non-Gaussian rotation
measure spread function, as explained in Section 1.2.2, a number of other effects
contribute to the difficulty of translating rotation measure cubes
into physical properties of the interstellar medium.

Firstly, in analogy to aperture synthesis, a limited range in
wavelength squared causes a limited sensitivity to large-scale Faraday
depth structures. In contrast with aperture synthesis, in RM synthesis
this can lead to a situation where the maximum detectable scale is
smaller than the Faraday depth resolution. Therefore, only
Faraday-thin components\footnote{Faraday-thin component is defined as
   a gaseous medium observed at a wavelength where the change of
   polarization angle through Faraday rotation is
   small. \citet{Brentjens05} define Faraday-thin as $\phi \lambda^2
   << 1$ and Faraday-thick as $\phi \lambda^2 >> 1$. A Faraday-thin
   component displays negligible internal Faraday depolarization.} and sharp gradients in
Faraday depth, such as the edges of Faraday-thick components, will
show up in a Faraday spectrum. The dependence on wavelength range of
the Faraday depth resolution $\delta\phi$, maximum detectable scale
$\Delta\phi_{max}$ and maximum detectable Faraday depth $\phi_{max}$
are given by \citet{Brentjens05} as

\begin{eqnarray}
\delta\phi &\approx& \frac{2\sqrt{3}}{\Delta\lambda^2}\\
\Delta\phi_{max} &\approx& \frac{\pi}{\lambda_{min}^2}\\
|\phi_{max}| &\approx& \frac{\sqrt{3}}{\delta\lambda^2}
\end{eqnarray}

Secondly, different Faraday depth features in a Faraday spectrum only
contain information about the amount of their Faraday depth, but not
necessarily about their distance. If along a line of sight Faraday
depth increases monotonically, i.e.\ if no magnetic field reversals
exist along the line of sight, then the distance order of Faraday
depth components is known. However, in the general ISM, with many
multi-scale magnetic field reversals, distance to Faraday components
is usually unknown. Only in exceptional cases, if one has
complementary rotation measures from a number of pulsars with known
distances along similar lines of sight, or if the Faraday depth
component has a counterpart with a known distance in another tracer,
is it possible to estimate the distance to a Faraday component.

Taking these caveats into account, a number of studies on RM synthesis
of diffuse Galactic synchrotron emission have been done, which show
mostly consistent results. \citet{Brentjens11} examines a $\sim
4^{\circ}\times7^{\circ}$ field around the Perseus galaxy cluster,
which mostly displays Galactic synchrotron emission, at a broad
frequency range around 350~MHz.  Synchrotron emitting
components are detected at multiple Faraday depths between $-50$ and
$+100$~rad~m$^{-2}$, are Faraday thin and spatially thin ($\leq 
40$~pc), and are well separated in Faraday depth space, suggesting that they
are flanked by Faraday-rotating-only parts of the ISM.

The same effect is noticed by \citet{Iacobelli13}, who study the
``ring stucture'' in the Fan region \citep[see
e.g.][]{Haverkorn03,Bernardi09} in RM synthesis around 150~MHz. They
also identify separate Faraday depth components, viz.\ the ring
structure and a foreground component which they associate with the
Local Bubble. Similarly, \citet{Pizzo10} notice in their Galactic
foreground studies in the direction of the galaxy cluster Abell~2255 at
multiple frequency bands from 150~MHz to 1200~MHz three
distinct ranges of Faraday depth with widely different morphologies.

These early RM synthesis studies of Galactic diffuse synchrotron
emission consistently conclude that the synchrotron emission is 
detected in discrete, often Faraday thin, structures with
widely different morphologies, interspersed with Faraday-rotating-only
components. It is tempting to interpret these observations as actual
small-scale variability in synchrotron emission in the ISM, or
discrete regions of excess emission. However, two other effects are at
play as well. Synchrotron emission dominates in locations where
$\vec{B} = \vec{B}_{\perp}$, while Faraday rotation only depends on
$B_{\parallel}$. This may cause the observed apparent anti-correlation
between synchrotron emission and Faraday rotation. Secondly, the
insensitivity of the technique to large Faraday-thick (emitting and Faraday rotating)
structures, which may mimic Faraday-thin emission components at the
edges of the Faraday depth range, will play a role.

Wavelet analysis can be successfully applied to Faraday depth cubes to
recognize magnetic features such as turbulence or large-scale magnetic
field reversals in nearby spiral galaxies or the intracluster medium
in galaxy clusters \citep{Beck12}. Low frequency data ($\sim 100$~MHz)
are needed to provide the necessary Faraday depth resolution, while
broad frequency coverage (up to several GHz) is crucial to make broad
Faraday structures detectable. In practice this requires combination
of broad-band data from various telescopes, such as the Global
Magneto-Ionic Medium Survey \citep[GMIMS,][]{Wolleben09}.

\subsection{Polarization gradients}

Linearly polarized intensity maps of diffuse synchrotron emission
consistently show narrow one-dimensional structures of complete
depolarization named {\em depolarization canals}
\citep{Haverkorn00}. Some of these depolarization canals are
observational artefacts due to missing short spacings in radio
interferometric observations, while other canals point to locations of
sharp jumps in rotation measure, i.e.\ sudden changes in electron
density and/or parallel magnetic field in the ISM \citep{Shukurov03,
  Haverkorn04b}. In addition, not all of these sudden changes in ISM
conditions are visible as depolarization canals.

The method of gradients in linear polarization was devised to obtain a
complete census of these locations of sudden change of conditions in
the ISM \citep{Gaensler11}. The vectorial polarization gradient is
calculated from the Stokes parameters $(Q, U)$ as

\begin{equation}
|\nabla\vec{P}| = \left\{ \left(\frac{\partial Q}{\partial x}\right)^2
  + \left(\frac{\partial Q}{\partial y}\right)^2 +
  \left(\frac{\partial U}{\partial x}\right)^2 + \left(\frac{\partial
      U}{\partial y}\right)^2\right\}^{1/2}
\label{e:polgrad}
\end{equation}

\begin{figure}[!h]
\includegraphics[width=\textwidth]{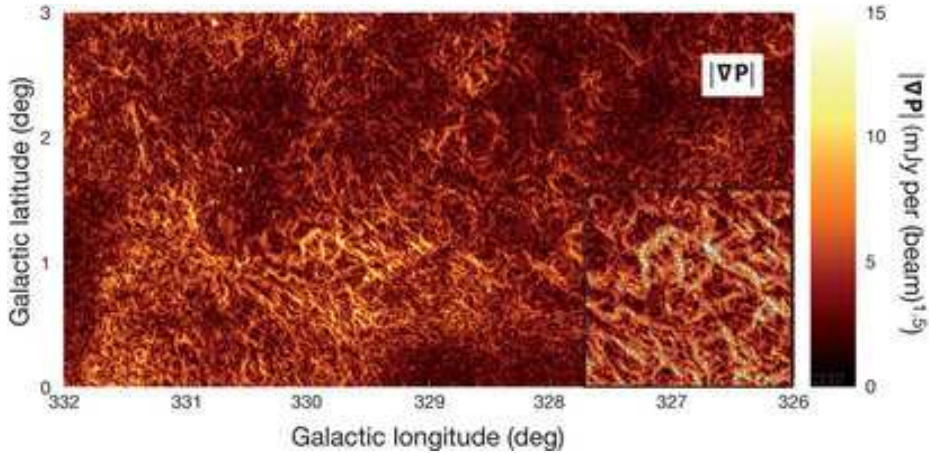}
\caption{The map of gradient of linear polarization $|\nabla\vec{P}|$
  as defined in equation~\ref{e:polgrad} for a field around the
  Galactic plane. The inset panel shows a
  $0.9^{\circ}\times0.9^{\circ}$ close-up of the brightest structure
  in $|\nabla\vec{P}|$, where the direction of the gradient is
  plotted only for strong ($> 5$~mJy~beam$^{1.5}$) gradient
  amplitudes. Figure reproduced from \citet{Gaensler11}.}
\label{f:polgrad}
\end{figure}

This gives a random-looking pattern of mostly one-dimensional
locations of high polarization gradient, of which depolarization
canals are a subset (see Fig.~\ref{f:polgrad}).

These polarization gradient filaments can be characterized by the
moments of the polarization gradient distribution. Simulations of
magnetohydrodynamic (MHD) turbulence show that the third and fourth
order moments (skewness $\gamma$ and kurtosis $\beta$) increase
monotonically with Mach number and depend on sonic Mach
number. Comparison of the observed values $\gamma = 0.3$ and $\beta =
0.9$ with simulated values for varying sonic Mach numbers indicates
that the magnetic turbulence in the ISM (at least in the field given
in Figure~\ref{f:polgrad}) is mildly subsonic to transonic
\citep{Gaensler11}. This is in agreement with estimates of the sonic
Mach number in the warm ionized medium from emission measure
distributions \citep{Hill08}.

Simulations also show that the filaments in polarization gradients can
be caused by either interacting shocks or random fluctuations in MHD
turbulence \citep{Burkhart12}. These authors also introduce the genus
method to characterize the polarization gradient maps. For subsonic
turbulence as in the ISM, where magnetic field fluctuations dominate
the polarization gradient topology, the topology is 'clumpy', as
opposed to supersonic turbulence which shows a ``Swiss cheese''
topology.

\section{Mysteries of Interstellar Turbulence}
In the paper to this point, we have reviewed the remarkable amount of information revealed by radioastronomical studies on turbulence in the ISM.  However, there remain a number of phenomena and effects that are not understood. In this section, we present what may be considered an agenda for future ISM radio propagation studies, that might clarify these issues.
We discuss topics in which additional measurements may contribute in a major way to advances in our understanding of interstellar turbulence, or cases in which emerging observational results appear difficult to understand, given our current vision of the interstellar medium and the turbulence in it.  These ``mysteries'' often involve input from the theory of plasma turbulence, and frequently rely on the latest results in that field.  In other cases, they represent known observational results of long standing that have eluded explanation.   

\subsection{The Existence of a Cascade in Interstellar Turbulence} 

Is there really a cascade in interstellar turbulence from the outer scale of 4 parsecs (or 100 pc) to the dissipation scale of order $100-500$~km?  Even in the case of well-studied solar wind turbulence this issue is not entirely resolved. Observations of solar wind turbulence indicate that it is comprised of Alfv\'{e}n waves propagating in both directions with respect to the large scale interplanetary magnetic field, i.e. towards and away from the Sun\footnote{A more precise and technical statement would be that both the positive and negative ``Elsasser Variables'' are present in solar wind turbulence.}.  Within the context of the most commonly discussed theories of solar wind turbulence, these counterpropagating waves are necessary for the existence of nonlinearities that produce the turbulent cascade.   In the case of the interstellar medium, observations are not adequate to demonstrate that counterpropagating Alfv\`{e}n waves are present, and it is not clear how such a discrimination could be done. In fact, it is not even clear that interstellar turbulence can be described as an ensemble of Alfv\'{e}n waves.

\subsection{Do We Understand the Flat RM Structure Functions?} 
The form of the rotation measure structure function for a turbulent, Faraday rotating medium with fluctuations in density $\delta n_e$ and magnetic field $\delta B_i$ ($B_i$ being a component of the magnetic field) was derived in \cite{Minter96}.  The expression presented there assumed vanishing correlation between $\delta n_e$ and all magnetic field components, but did not exclude the possibility of a correlation between the density and the magnitude of the field.  The expression of \cite{Minter96} agrees with that of \cite{Simonetti88} in the limiting case of density fluctuations in a uniform magnetic field. 
 
Rotation Measure structure functions should have a logarithmic slope of 5/3 if both density and magnetic field fluctuations have a Kolmogorov spectrum, and an observationally indistinguishable value of 3/2 if a Kraichnan spectrum applies.  This results holds if the two lines of sight are separated by a distance which is in the inertial subrange of the turbulence.  Such a slope is rarely measured; a number of independent investigations have found that the RM structure functions on angular lags of several tenths of a degree to several degrees have logarithmic slopes of $\sim 1/2$, or even flatter.

The interpretation of this result has been that the angular lags probed correspond to large scales in the interstellar medium, of order the outer scale or larger \citep{Minter96,Haverkorn06,Haverkorn08}.  \cite{Minter96} suggest that the aforementioned data are probing 2D turbulence that exists in sheets, and that the fully 3D component of the turbulence is on spatial scales less than the thickness of these sheets, with corresponding angular scales of order a few tenths of a degree or less. These analyses are, in fact, the basis of the claim that the outer scale is of order 1 - 5 parsecs in extent.  

Nonetheless, it would be comforting to actually measure, in a clear and unambiguous fashion, the transition from a $\sim 1/2$ logarithmic slope on angular lags $\geq 1^{\circ}$ to $\sim 5/3$ on angular lags $\leq 0.2^{\circ}$.  This would securely establish the value of the outer scale in the WIM turbulence.  Without such measurements, we will continue to be tormented by the specter of an ISM in which we {\em are} measuring the inertial range of the turbulence (assuming this to be a meaningful concept), and that we lack an explanation for its logarithmic slope. Such a flat slope for an inertial subrange of interstellar turbulence would constitute a major paradox for our understanding of interstellar turbulence.  A flat spectrum (power law index $\sim \frac{1}{2}$ instead of $\simeq \frac{5}{3}$ would predict density fluctuations on the scales responsible for radio wave scattering (see Section 4.3) that are far too large to be compatible with the observed magnitude of radio scintillations of pulsars and extragalactic radio sources.  
 
\subsection{What is the Significance of the ``Pulsar Arcs''} One of the most intriguing developments of the last decade in the study of interstellar turbulence has been the discovery of ``pulsar arcs'' \citep{Walker04,Cordes06}. This phenomenon seems most easily explicable if the turbulence responsible for interstellar turbulence is confined to one, or at most, a few thin sheets. A recent overview of the observational properties of the arcs and their interpretation is given in \cite{Rickett11}. 

The existence of the arcs is thus linked to one of the most intriguing ``mysteries'' of ISM turbulence, i.e. whether that turbulence is widely distributed through the Galaxy, or confined to spatially restricted and widely separated regions of intense turbulence. The resolution of this matter has obvious import for our understanding of the mechanisms which generate the turbulence.  

As noted above, the existence of the arcs for many pulsars seems to suggest that the turbulence exists in thin sheets.  However, the research to definitely prove this has not yet been done.  \cite{Cordes06} point to the desirability of calculations that would investigate the properties, including the existence of arcs caused by extended turbulent media.  

Other types of radio scattering measurements can also address the question of the distribution of the turbulence.  A particularly promising approach was investigated by \cite{Gwinn93} who compared angular broadening and pulse broadening measurements for a sample of 10 pulsars.  As discussed in \cite{Gwinn93}, the angular width and the temporal width of pulse broadening have different dependences on the distribution of turbulent plasma along the line of sight.  In principle, a comparison between pulse broadening and angular broadening can distinguish between uniformly distributed turbulence and that concentrated in a thin screen.  \cite{Gwinn93} concluded that their data were consistent with a uniform distribution of turbulence, except for pulsars such as the Crab Nebula and Vela Nebula pulsars, for which the turbulence is partially contained in a screen that is naturally associated with a supernova remnant.  

Another investigation into this matter \citep{Bhat04} used only pulse broadening measurements.  These authors utilized the fact that the shape of the pulse broadening function is different for screens and uniform, extended media.  The goal of the analysis of \cite{Bhat04} was to determine if screen or extended media better fit the observations of a sample of 98 pulsars.  They found that some pulsars in their sample were better fit by extended media, and others by thin screens.  This study would therefore indicate that there is no general rule regarding the distribution of turbulence in interstellar space.  

A final relevant study is that of \cite{Linsky08}, who convincingly associated the turbulence responsible for intraday flux variations  of two quasars with a region of interaction between two of the clouds in the Very Local Interstellar Medium (VLISM). In this specific case, the radio wave scattering is dominated by turbulence in a relatively thin region of interaction between two independent media.  However, it is not clear if these conclusions for the VLISM are applicable to the general ISM.  

In summary, at the present time observations are ambiguous as to whether the small scale turbulence which is responsible for radio wave scintillations, and which is a focus of attention in this paper, uniformly fills the ISM, is confined to thin layers on presumptive interfaces in the ISM, or is a combination of these two limiting cases.  Different observational studies reach different conclusions.  Future research could improve our state of knowledge.  Work in the theory of scintillations could indicate if thin turbulent screens are necessary for the existence of pulsar arcs, or if these features can also arise in extended, turbulent media.  Equally promising would be a new investigation along the lines of \cite{Gwinn93}, utilizing angular broadening measurements made with the Very Long Baseline Array (VLBA), an instrument which has greater sensitivity and accuracy than the instrument used in \cite{Gwinn93}, and pulse broadening analyses as in \cite{Bhat04}.   

\subsection{What Generates Interstellar Turbulence?}  
Plasma turbulence, as revealed by the density fluctuations responsible for radio scintillations, appears to be very widely distributed in the interstellar medium. There are also indicators of turbulence in the neutral gas, such as spectral line widths enhanced over their thermal values.  Evidence exists for turbulence in most, or all of the phases of the ISM listed in Table 1, and this turbulence exists on a wide range of spatial scales.  A current review of interstellar turbulence in general is \cite{Elmegreen04}.  
 The question then arises as to the mechanism responsible for its generation. A general consensus holds that the free energy source is expanding supernova remnants, stellar superbubbles, and expanding HII regions, as well as magnetorotational, shear, or other instabilities associated with Galactic rotation \citep{Norman96,Elmegreen04,MacLowKlessen04,Hill12}. The small scale fluctuations that we detect in radio scintillations might be generated by baroclinic effects\footnote{Baroclinic effects involve the generation of fluid vorticity by misaligned gradients of pressure and density.} at the expanding interfaces between supernova remnants, stellar bubbles, and the ISM. There is also the most obvious possibility, in which these small scale irregularities arise as a consequence of a cascade from large injection scales.  This suggestion encounters the difficulty that known dissipation mechanisms are not restricted to small spatial scales (see Section 6.6 below).  

A number of studies have estimated the volumetric power input to interstellar turbulence from a variety of astronomical sources; these are summarized in \cite{Elmegreen04}.  The relative contributions of these sources are estimated by \cite{Norman96}.  \cite{Elmegreen04} quote a global turbulent input power density of $3 \times 10^{-26}$ ergs/sec/cm$^3$, attributing this theoretical estimate to \cite{MacLowKlessen04}.  Interesting, and perhaps fortuitously, this turbulent power density is very close in magnitude to both the estimated volumetric heating rate in the WIM from the dissipation of turbulence due to ion-neutral collisions, and the cooling rate of the WIM \citep{Minter97,Spangler03}.  These sums are then consistent, though not uniquely so, with a picture in which turbulent energy is input by supernova and stellar associations on the scale of parsecs or tens of parsecs, cascades down to scales comparable to and smaller than the ion-neutral collisional scale, where it is dissipated and then radiated away by the glow of the WIM. \cite{Elmegreen04} estimate that the power input from Galactic rotation is significantly smaller than the numbers above. \cite{Elmegreen04} also make the important point that turbulent input power densities seems to be significantly higher in the denser parts of the ISM, suggesting distinct and segregated turbulent generation mechanisms in different phases of the ISM.  

A final point about generation of turbulence is that there may be a problem with the distribution or diffusion of turbulence throughout the ISM. The preceding discussion assumes that all of the processes involved in turbulent power input, i.e. generation at the outer scale, cascade through wavenumber, and dissipation of small spatial scales, are spatially co-located. However, supernova remnants and stellar bubbles occupy a very small fraction of the ISM, and turbulent damping limits the extent to which turbulence can propagate from the generation site to locations throughout the WIM \citep{Spangler07,Spangler11b}. Observations of scintillations of pulsars and extragalactic radio sources, on the other hand, indicate that turbulence is widely distributed through at least the WIM phase of the ISM.  

However, it must be admitted that the possible difficulty raised in \cite{Spangler07,Spangler11b} has not generated ``weeping and the gnashing of teeth'' in the interested community.  The response from that community has been that the role of supernova remnants is to excite a global Galactic system of flows, which then generate turbulence throughout the Galaxy via velocity shear.  An evaluation of this matter will depend on a better understanding of whether ISM turbulence is produced by a limited number of point sources, or by processes such as shear that occur throughout the Galaxy.   

\subsection{Removal of Fast Magnetosonic Waves from Interstellar Turbulence} Fast Magnetosonic waves are one of the three MHD wave modes, so one would expect them to comprise part of interstellar turbulence. \cite{Cho02,Cho03} and \cite{Klein12} have formally investigated the partition of MHD turbulence into fluctuations possessing the properties of these modes, as well as the generation of slow mode and fast mode-like fluctuations from predominantly Alfv\'{e}nic turbulence.  However, it has been argued on observational grounds that Fast Mode waves can only constitute a negligibly small fraction of the energy in interstellar turbulence.  The argument is based on the rapid damping of Fast Mode waves on thermal ions for conditions appropriate to the Warm Ionized Medium.  If a sizeable fraction of the energy in interstellar turbulence is in the form of Fast Mode waves, then the large power input to the interstellar medium would exceed the cooling capacity of the WIM gas \citep{Spangler91,Spangler03}. Interestingly enough, these waves also seem to be absent from the solar corona and the solar wind at 1 a.u.. \cite{Harmon05} make a convincing argument that a substantial contribution of Fast Mode waves to the coronal turbulence budget is incompatible with spaced-receiver propagation measurements. \cite{Klein12} also argue that Fast Modes waves, or fluctuations possessing Fast Mode properties, constitute an insignificant portion of solar wind turbulence at 1 a.u.. The analysis of \cite{Klein12} is based on calculations of simulated turbulence, consisting of a superposition of Fast Mode waves, Slow Mode waves, and Alfv\'{e}n waves.  These simulated realizations of turbulence are compared with actual measurements of solar wind turbulence, especially the density-magnetic field correlation function.  \cite{Klein12} find that the realizations that resemble the true, observed turbulence are those with an insignificant fraction of Fast Mode waves. It should be mentioned before leaving this topic that the absence of Fast Mode waves in heliospheric plasmas is a characteristic of plasmas far from shocks or other sources of unstable particle distributions.  Shocks produce ion streaming instabilities which in turn generate beautiful, large amplitude Fast Magnetosonic waves, the best known examples of which are the waves upstream of the Earth's bow shock \citep[see][for an entry point to a large literature]{Hoppe81}.  However, it seems to be the case that these Fast Mode waves are confined to relatively thin layers that bound strong shocks in the solar wind.  To conclude this subsection, whether such a  minor role for the Fast Mode in astrophysical turbulence is due to enhanced damping, or the turbulence generation mechanisms remains to be determined by future research.  

\subsection{Do We Understand the Lack of a Spectral Break at the Ion-Neutral Collisional Scale?} As discussed in Section 3.1, there is observational evidence for a power law spectrum of interstellar turbulence from scales of at least a few parsecs, down to scales as small as 100 km.  The existence of a power law spectrum seems to indicate, on general grounds of dimensional analysis, that there are no fundamental scales between the outer scale on which stirring is done, and the inner scale where dissipation occurs.  This assumption is in stark contrast to the situation for the WIM phase of the interstellar medium, in which a fundamentally-defined scale, the collisional scale $l_c \equiv \frac{V_A}{\nu_{in}}$, where $V_A$ is the Alfv\'{e}n speed and $\nu_{in}$ is the ion-neutral collisional scale, lies between the inner and outer scales ($l_c \simeq 10^{15} - 10^{16}$ cm). In the WIM the neutral atoms are helium, which is partially or fully neutral. This oddity was discussed by \cite{Armstrong95}.

\citet[][this paper also contains references to earlier results by these authors]{Cho03} and \cite{Oishi06} present numerical simulations showing that irregularities exist on scales smaller than the ion-neutral collisional scale, and conclude that ion-neutral collisional effects do not truncate the turbulent cascade. A different conclusion apears to be reached by \cite{Shaikh08}. \cite{Shaikh08} find that while the turbulent cascade continues at wavenumbers larger than that corresponding to the ion-neutral collisional scale, the fundamental physics of the nonlinear interaction is modified.  These authors claim that the magnetic field and velocity power spectra of the ionized fluid are significantly steepened in comparison with the same spectra of a fully-ionized plasma.  This result would seem to be discordant with the observed Kolmogorov spectrum of density fluctuations down to much smaller spatial scales. Further discussion of this interesting question should include recognition of the intrinsic anisotropy of MHD turbulence (see Section 3.4).  For Alfv\'{e}nic turbulence, the ion-neutral collisional interaction is determined by the parallel wavenumber $k_{\parallel}$, whereas the irregularities responsible for radio wave scintillation almost certainly have $\frac{k_{\perp}}{k_{\parallel}} \gg 1$. Whether the anisotropy of turbulence, which will be large at these scales, is enough to resolve this question remains to be determined.  

\subsection{The Outer Scale of Interstellar Turbulence and Cosmic Ray Propagation} 
In Section 3.6 above, we noted that several independent investigations find that the outer scale of turbulence in the WIM must be of order a few parsecs.  There is an associated curiosity that was raised in \cite{Spangler01}, but does not appear to have been discussed since.  If there is a break in the turbulence corresponding to an outer scale of 4 parsecs, then there should be an associated change in the transport properties of cosmic rays which are resonant with such irregularities, i.e. those with energies of $10^{15}- 10^{16}$ eV.  In fact, the famous ``knee'' in the cosmic ray spectrum occurs here, but other mechanisms are normally invoked for its existence.  Should the change in turbulence properties at scales that resonate with such cosmic rays be added to the mechanisms considered? In considering this matter, it should be recognized that the outer scale of turbulence in the Galactic halo is probably large, of order 100 pc (see Section 3.6 above).  The larger fluctuations in the Galactic halo could resonate with higher energy cosmic rays than the smaller fluctuations in the WIM of the spiral arms.  The fluctuations in the halo may dominate the Galactic transport of cosmic rays.  

\subsection{Can Observations Provide a Connection to Theories of Kinetic Processes in Turbulence?} One of the most intriguing recent developments in the study of plasma turbulence has been elucidation of the role of kinetic processes in turbulence (i.e.\ those described by the Vlasov equation rather than MHD), and the observational support for these ideas in spacecraft measurements of solar wind turbulence (Section 3.5 above). Can we find similar evidence of kinetic processes in the interstellar medium?  Do the same kinetic processes which appear crucial in the solar wind, such as Landau damping of kinetic Alfv\'{e}n waves, play an important role in the interstellar medium? In kinetic damping processes, energy will flow to either ions or electrons, depending on which is resonant with the fluctuations being damped.  However, observations of the WIM plasma, the best diagnosed astrophysical plasma, show temperature equilibration between electrons and different ion species \citep{Haffner09}.  A similar situation occurs in the clouds of the Very Local Interstellar Medium, where the same temperature characterizes neutral atoms as well as several ions with different masses (and therefore cyclotron frequencies, \citep{Spangler11b}).

\subsection{Why is the Spectrum of Plasma Turbulence the Same in HII Regions, the Warm Ionized Medium, and the Solar Wind?} 
The \cite{Armstrong95} result of a Kolmogorov density fluctuation spectrum over several decades pertains to the WIM component of the interstellar medium, having been established from observations of relatively nearby pulsars and extragalactic radio sources whose lines of sight are at high galactic latitudes.  Radio wave propagation measurements made on heavily-scattered lines of sight that pass through HII regions \citep[e.g.][]{Spangler88,Molnar95,Rickett09} are also consistent with a Kolmogorov density spectrum. Finally, the plasma of the solar corona and inner solar wind also has a Kolmogorov spectrum, particularly in the slow solar wind. In the solar wind, the spectral of magnetic field and flow velocity are observed to possess inertial subranges with spectra that are close to Kolmogorov.  Whether the spectra are both exactly Kolmogorov, both slightly flatter than Kolmogorov, or different for the two fields remains a point of contention.  \cite{Boldyrev11} claim that there is a slight difference in power law indices of the magnetic and velocity spectra, but this has been disputed by \cite{Beresnyak10}.  Regardless of the resolution of these important matters dealing with the physical nature of plasma turbulence, it remains empirically the case that density, magnetic field, and velocity have ``Kolmogorov-like'' spectra in the solar wind, and plasmas in the WIM and HII regions have similar density spectra.  This result is, perhaps, somewhat unexpected since the mechanisms for generation of the turbulence at the outer scale are presumably quite different in these different media, as might be the plasma $\beta$ that determines the dissipation mechanisms. Is the similarity of the turbulence in these quite different media a consequence of the universality of turbulence?  

\section{Summary and Conclusions}
Radio propagation observations yield a surprising amount of quantitative information about the plasma state of the interstellar medium, particularly for turbulence in the WIM phase of the ISM and HII regions.  The measurements emphasized in this paper have been Faraday rotation of linearly polarized signals that have propagated through the ISM during the passage from extragalactic radio sources to the Earth, frequency-dependent polarization characteristics of the Galactic synchrotron radiation, and scintillations of Galactic and extragalactic radio sources due to small scale density fluctuations in the ISM.  From this information, we can deduce the amplitude and spectral properties of interstellar turbulence.  We have information on the outer and inner scales of this turbulence; the corresponding inertial subrange extends over roughly 10 decades.  In some respects, interstellar turbulence resembles the extensively studied turbulence in the solar wind.  However, there appear to be significant differences as well.  In spite of impressive progress in this field, there are several (at least) poorly-understood aspects of interstellar turbulence that warrant the term ``mysteries''.  Several of these aspects are discussed in Section 6 above.  Some of these could be addressed in a significant way with new observations on new or substantially upgraded radio telescopes such as the VLBA and LOFAR.

\begin{acknowledgements} This work was supported at the University of Iowa by grants AST09-07911 and ATM09-56901 from the National Science Foundation of the United States. M.H. acknowledges the support of research programme 639.042.915, which is partly financed by the Netherlands Organisation for Scientific Research (NWO). The authors acknowledge the work of Jacob J. Buffo of the University of Iowa in the analysis of model structure functions, contained in Figures 4 and 5. We thank James Cordes of Cornell University for providing the beautiful pulsar dynamic spectrum in Figure~\ref{fig:dynamic_spectrum} as an illustration of one of the phenomena of radio wave propagation in a random medium. We also appreciate the advice and recommendations of Dr. Cordes on the question of the spatial uniformity of interstellar turbulence.  Finally, we thank Olga Alexandrova, Rainer Beck, and Alex Lazarian for interesting, helpful, and collegial readings of this paper.     \end{acknowledgements}

\bibliographystyle{aps-nameyear}      
\bibliography{example}   
\nocite{*}


\end{document}